# Generating Extractive Summaries of Scientific Paradigms

**Vahed Qazvinian**　　　　　　　　　　　　　　　　　　　VAHED@UMICH.EDU
*Department of EECS,*
*University of Michigan, Ann Arbor, MI, 48109*

**Dragomir R. Radev**　　　　　　　　　　　　　　　　　　RADEV@UMICH.EDU
*Department of EECS & School of Information,*
*University of Michigan, Ann Arbor, MI, 48109*

**Saif M. Mohammad**　　　　　　　　SAIF.MOHAMMAD@NRC-CNRC.GC.CA
*National Research Council Canada,*
*Ottawa, Ontario, Canada, K1A 0R6*

**Bonnie Dorr**　　　　　　　　　　　　　　　　　　BONNIE@UMIACS.UMD.EDU
**David Zajic**　　　　　　　　　　　　　　　　　　DMZAJIC@UMIACS.UMD.EDU
**Michael Whidby**　　　　　　　　　　　　　　　　　MAWHIDBY@UMD.EDU
**Taesun Moon**　　　　　　　　　　　　　　　　　　TSMOON@UMD.EDU
*Institute for Advanced Computer Studies & Computer Science,*
*University of Maryland, College Park, MD, 20742*

## Abstract

Researchers and scientists increasingly find themselves in the position of having to quickly understand large amounts of technical material. Our goal is to effectively serve this need by using bibliometric text mining and summarization techniques to generate summaries of scientific literature. We show how we can use citations to produce automatically generated, readily consumable, technical extractive summaries. We first propose C-LexRank, a model for summarizing single scientific articles based on citations, which employs community detection and extracts salient information-rich sentences. Next, we further extend our experiments to summarize a set of papers, which cover the same scientific topic. We generate extractive summaries of a set of Question Answering (QA) and Dependency Parsing (DP) papers, their abstracts, and their citation sentences and show that citations have unique information amenable to creating a summary.

## 1. Introduction

In today's rapidly expanding disciplines, scientists and scholars are constantly faced with the daunting task of keeping up with knowledge in their field. In addition, the increasingly interconnected nature of real-world tasks often requires experts in one discipline to rapidly learn about other areas in a short amount of time. Cross-disciplinary research requires scientists in areas such as linguistics, biology, and sociology to learn about computational approaches and applications such as computational linguistics, biological modeling, and social networks. Authors of journal articles and books must write accurate summaries of previous work, ranging from short summaries of related research to in-depth historical notes. Interdisciplinary review panels are often called upon to review proposals in a wide range of areas, some of which may be unfamiliar to panelists. Thus, they must learn about a new discipline "on the fly" in order to relate their own expertise to the proposal.

Our goal is to effectively serve these needs by combining two currently available technologies: (1) bibliometric lexical link mining that exploits the structure of citations and (2) summarization techniques that exploit the content of the material in both the citing and cited papers.

It is generally agreed upon that manually written abstracts are good summaries of individual papers. More recently, Qazvinian and Radev (2008) argued that *citation sentences* (i.e., set of sentences that appear in other papers and cite a given article) are useful in creating a summary of important contributions of a research paper. Kaplan, Iida, and Tokunaga (2009) introduced "citation-site" as a block of text that includes a citation and discusses the cited text. This work used a machine learning method for extracting citations from research papers and evaluates the result using an annotated corpus of 38 papers citing 4 articles. Moreover, Qazvinian and Radev (2010) showed the usefulness of using implicit citations (i.e., *context sentences*, sentences that occur before or after a citation sentence and do not explicitly cite the target paper, but discuss its contributions) in summary generation. Teufel (2005) argued that citations could contain subjective content, and that this content can be exploited for summary generation. Additional work (Mohammad et al., 2009) demonstrated the usefulness of citations for producing multi-document summaries of scientific articles. Follow-up work indicated that further improvements to citation handling enables the production of more fluent summaries (Whidby, 2012).

In our work, we develop summarization systems that exploit citations. Specifically,

- We compare and contrast the usefulness of abstracts and of citations in automatically generating a technical summary on a given topic from multiple research papers. Our findings suggest that abstracts and citations have some overlapping information but they also have a significant amount of unique summary-amenable information. Particularly, we provide evidence that citation sentences contain crucial information that is not available, or hard to extract, from abstracts and papers alone.

- We propose C-LexRank, a graph based summarization system. This method models a set of citing sentences as a network in which vertices are sentences and edges represent their lexical similarity. C-LexRank then identifies vertex communities (clusters) in this network, and selects sentences from different communities to increase diversity in the summary. Using 30 different sets of citation sentences extracted from 6 different NLP topics in the ACL[1] Anthology Network, we show that C-LexRank is effective in producing a summary of a paper's contributions. We compare C-LexRank with a wide range of state-of-the-art summarization systems that leverage diversity (MMR, DivRank, MASCS), employ graph structure (DivRank, LexRank), or employ sentence compression (MASCS) to produce a summary.

- We extend our experiments from summarizing the contributions of a single article to generating summaries of scientific topics. Our evaluation experiments for extractive summary generation are applied to a set of 10 papers in the research area of Question Answering (QA) and another set of 16 papers on Dependency Parsing (DP).

---

1. Association for Computational Linguistics

We provide some background for this work including the primary features of a technical summary and also the types of input that are used in our study (full papers, abstracts, and citation sentences).

## 1.1 Background

Automatically creating technical extractive summaries is significantly distinct from traditional multi-document summarization. Below we describe the primary characteristics of a technical extractive summary and we present different types of input texts that we used for the production of extractive summaries.

### 1.1.1 TECHNICAL EXTRACTIVE SUMMARIES

In the case of multi-document summarization, the goal is to produce a readable presentation of multiple documents, whereas in the case of technical summary creation, the goal is to convey the key features and basic underpinnings of a particular field, early and late developments, important contributions and findings, contradicting positions that may reverse trends or start new sub-fields, and basic definitions and examples that enable rapid understanding of a field by non-experts.

A prototypical example of a technical summary is that of "chapter notes," i.e., short (50–500 word) descriptions of sub-areas found at the end of chapters of textbooks, such as Jurafsky and Martin's (2008). One might imagine producing such descriptions automatically, then hand-editing them and refining them for use in an actual textbook.

Previously Mohammad et al. (2009) conducted a human analysis of these chapter notes and revealed a set of conventions, an outline of which is provided here (with example sentences in italics):

1. Introductory/opening statement: *The earliest computational use of X was in Y, considered by many to be the foundational work in this area.*

2. Definitional follow up: *X is defined as Y.*

3. Elaboration of definition (e.g., with an example): *Most early algorithms were based on Z.*

4. Deeper elaboration, e.g., pointing out issues with initial approaches: *Unfortunately, this model seems to be wrong.*

5. Contrasting definition: *Algorithms since then...*

6. Introduction of additional specific instances / historical background with citations: *Two classic approaches are described in Q.*

7. References to other summaries: *R provides a comprehensive guide to the details behind X.*

The notion of *text level categories* or *zoning* of technical papers—related to the summary components enumerated above—has been investigated previously in the work of Teufel and Moens (2002) and Nanba, Kando, and Okumura (2000). These earlier works focused on the *analysis* of scientific papers based on their rhetorical structure and on determining the portions of papers that contain new results, comparisons to earlier work, etc. The work described here focuses on the *synthesis* of technical summary based on knowledge gleaned from rhetorical structure not unlike that of the work of these earlier researchers, but guided by structural patterns along the lines of the conventions listed above.

Although our current approach to summary creation does not yet incorporate a fully pattern-based component, our ultimate objective is to apply these patterns to guide the creation and refinement of the final output. As a first step toward this goal, we use citation sentences (closest in structure to the patterns identified by convention 7 above) to pick out the most important content for summary creation.

### 1.1.2 SCHOLARLY TEXTS

Published research on a particular topic can be summarized from two different kinds of sources: (1) where an author describes her own work and (2) where others describe an author's work (usually in relation to their own work). The author's description of her own work can be found in her paper. How others perceive her work is spread across other papers that cite her work.

Traditionally, technical summary generation has been tackled by summarizing a set of research papers pertaining to the topic. However, individual research papers usually come with manually-created "summaries"—their abstracts. The abstract of a paper may have sentences that set the context, state the problem statement, mention how the problem is approached, and the bottom-line results—all in 200 to 500 words. Thus, using only the abstracts (instead of full papers) as input to a summarization system is worth exploring.

Whereas the abstract of a paper presents what the authors think to be the important aspects of a paper, the citations to a paper capture what others in the field perceive as the broader contributions of the paper. The two perspectives are expected to have some overlap in their content, but the citations also contain additional information not found in abstracts (Elkiss, Shen, Fader, Erkan, States, & Radev, 2008; Nakov & Hearst, 2012). For example, authors may describe how a particular methodology from one paper was combined with another from a different paper to overcome some of the drawbacks of each. Citations are also indicators of what contributions described in a paper were influential over time.

Another feature that distinguishes citations from abstracts is that citations tend to have a certain amount of redundant information. This is because multiple papers may describe the same contributions of a target paper. This redundancy can be exploited by automatic systems to determine the important contributions of the target paper.

Our goal is to test the hypothesis that an effective technical summary will reflect information on research not only from the perspective of its authors but also from the perspective of others who use, commend, discredit, or add to it. Before describing our experiments with technical papers, abstracts, and citations, we first summarize relevant prior work that used these sources of information as input.

The rest of this paper is organized as follows. After reviewing the related work, we present an analysis of citations and demonstrate that they contain summary-amenable information. In the process, we develop C-LexRank, a citation-based summarization system. In Section 5, we show that state-of-the-art automatic summarization systems create more contentful summaries of citations of individual documents than those created simply by random sampling. We also show that C-LexRank performs better than other state-of-the-art summarization systems when producing both 100- and 200-word extracts. In Section 6, we extend our experiments to summarize a set of papers representing the same scientific topic using the source texts as well as citations to the topic papers. Additionally, we show the usefulness of citation sentences in automatically generating a technical summary on a given topic. We observe that, as expected, abstracts are useful in summary creation, but, notably, we also conclude that citations contain crucial information not present in (or at least, not easily extractable from) abstracts. We further discover that abstracts are author-biased and thus complementary to the broader perspective inherent in citation sentences; these differences enable the use of a range of different levels and types of information in the summary.

## 2. Related Work

In this section, we review related prior work in two categories. First, we review previous research on citation analysis, and then we discuss prior work on capturing diversity in automatic text summarization.

### 2.1 Citation Analysis

Previous work has analyzed citation and collaboration networks (Teufel, Siddharthan, & Tidhar, 2006; Newman, 2001) and scientific article summarization (Teufel & Moens, 2002). Bradshaw (2002, 2003) benefited from citations to determine the content of articles and introduce "Reference Directed Indexing" to improve the results of a search engine. Nanba, Abekawa, Okumura, and Saito (2004) and Nanba et al. (2000) analyzed citation sentences and automatically categorize citations into three groups using 160 pre-defined phrase-based rules. This categorization was then used to build a tool to help researchers analyze citations and write scientific summaries. Nanba and Okumura (1999) also discussed the same citation categorization to support a system for writing a survey. Nanba and Okumura (1999) and Nanba et al. (2000) reported that co-citation implies similarity by showing that the textual similarity of co-cited papers is proportional to the proximity of their citations in the citing article.

Previous work has shown the importance of the citation sentences in understanding scientific contributions. Elkiss et al. (2008) performed a large-scale study on citations and their importance. They conducted several experiments on a set of $2,497$ articles from the free PubMed Central (PMC) repository[2] and 66 from ACM digital library. Results from this experiment confirmed that the average cosine between sentences in the set of citations to an article is consistently higher than that of its abstract. They also reported that this number is much greater than the average cosine between citation sentences and a randomly chosen document, as well as between citation sentences and the abstract. Finally, they concluded that the content of citing sentences has much greater uniformity than the content of the corresponding abstract, implying that citations are more focused and contain additional information that does not appear in abstracts.

---

2. http://www.pubmedcentral.gov

Nakov and Hearst (2012) performed a detailed manual study of citations in the area of molecular interactions and found that the set of citations to a given target paper cover most information found in the abstract of that article, as well as 20% more concepts, mainly related to experimental procedures.

Kupiec, Pedersen, and Chen (1995) used the abstracts of scientific articles as a target summary. They used 188 Engineering Information summaries that are mostly indicative in nature. Kan, Klavans, and McKeown (2002) used annotated bibliographies to cover certain aspects of summarization and suggest guidelines that summaries should also include metadata and critical document features as well as the prominent content-based features.

Siddharthan and Teufel (2007) described a new reference task and show high human agreement as well as an improvement in the performance of *argumentative zoning* (Teufel, 2005). In argumentative zoning—a rhetorical classification task—seven classes (Own, Other, Background, Textual, Aim, Basis, and Contrast) are used to label sentences according to their role in the author's argument.

The problem of automatic related work summarization is addressed by Hoang and Kan (2010). In their work, Hoang and Kan used a set of keywords representing a hierarchy of paper topics and assigned a score to each input sentence to construct an extractive summary.

Athar (2011) addressed the problem of identifying positive and negative sentiment polarity in citations to scientific papers. Similarly, Athar and Teufel (2012) used context-enriched citations to classify scientific sentiment towards a target paper.

## 2.2 Leveraging Diversity in Summarization

In summarization, a number of previous methods have focused on the diversity of perspectives. Mei, Guo, and Radev (2010) introduced DivRank, a diversity-focused ranking methodology based on reinforced random walks in information networks. Their random walk model, which incorporates the rich-gets-richer mechanism to PageRank with reinforcements on transition probabilities between vertices, showed promising results on the Document Understanding Conference (DUC) 2004 dataset. DivRank is a state-of-the-art graph-based method and it leverages the diversity of perspectives in summarization. Therefore, we chose this algorithm as an important baseline in our experiments and we will discuss it in more detail in Section 4.

A similar ranking algorithm, described by Zhu, Goldberg, Van Gael, and Andrzejewski (2007), is the *Grasshopper* ranking model, which leverages an absorbing random walk. This model starts with a regular time-homogeneous random walk, and in each step the vertex with the highest weight is set as an absorbing state. Paul, Zhai, and Girju (2010) addressed the problem of summarizing opinionated text using *Comparative LexRank*, a random walk model inspired by LexRank (Erkan & Radev, 2004). Comparative LexRank first assigns different sentences to clusters based on their contrastiveness with each other. It then modifies the graph based on cluster information and performs LexRank on the modified cosine similarity graph.

Perhaps the most well-known summarization method to address diversity in summarization is Maximal Marginal Relevance (MMR) (Carbonell & Goldstein, 1998). This method is based on a greedy algorithm that selects sentences in each step that are the least similar to the summary so far. We compare our summarization output with that of MMR and discuss this algorithm in more details in Section 4.

In prior work on evaluating independent contributions in content generation, Voorhees (1998) studied IR systems and showed that relevance judgments differ significantly between humans but relative rankings show high degrees of stability across annotators. In other work, van Halteren and Teufel (2004) asked 40 Dutch students and 10 NLP researchers to summarize a BBC news report, resulting in 50 different summaries. They also used 6 DUC-provided summaries, and annotations from 10 student participants and 4 additional researchers, to create 20 summaries for another news article in the DUC datasets. They calculated the Kappa statistic (Carletta, 1996; Krippendorff, 1980) and observed high agreement, indicating that the task of atomic semantic unit (factoid) extraction can be robustly performed in naturally occurring text, without any copy-editing.

The diversity of perspectives and the growth of the factoid inventory (Qazvinian & Radev, 2011b) also affects evaluation in text summarization. Evaluation methods are either extrinsic, in which the summaries are evaluated based on their quality in performing a specific task (Spärck-Jones, 1999) or intrinsic where the quality of the summary itself is evaluated, regardless of any applied task (van Halteren & Teufel, 2003; Nenkova & Passonneau, 2004). These evaluation methods assess the information content in the summaries that are generated automatically.

## 3. Citation-Based Summarization

The ACL Anthology Network[3] (AAN) is a manually curated resource built on top of the ACL Anthology[4] (Bird, Dale, Dorr, Gibson, Joseph, Kan, Lee, Powley, Radev, & Tan, 2008). AAN includes all the papers published by ACL and related organizations as well as the Computational Linguistics journal over a period of four decades. AAN consists of more than $18,000$ papers from more than $14,000$ authors, each distinguished with a unique ACL ID, together with their full-texts, abstracts, and citation information. It also includes other valuable metadata such as author affiliations, citation and collaboration networks, and various centrality measures (Radev, Muthukrishnan, & Qazvinian, 2009; Joseph & Radev, 2007).

To study citations across different areas within Computational Linguistics, we first extracted six different sets of papers from AAN corresponding to 6 different NLP topics: Dependency Parsing (DP), Phrase-based Machine Translation (PBMT), Text Summarization (Summ), Question Answering (QA), Textual Entailment (TE), and Conditional Random Fields (CRF). To build each set, we matched the topic phrase against the title and the content of AAN papers, and extracted the 5 highest cited papers. Table 1 shows the number of articles and the number of citation sentences in each topic[5]. The number of citations in each set shows that number of sentences that are used as an input to various summarization systems in our experiments.

---

3. http://clair.si.umich.edu/anthology/
4. http://www.aclweb.org/anthology-new/
5. The number of incoming citations are from AAN's 2008 release.

| | ACL ID | Title | Year | # citations |
|---|---|---|---|---|
| **DP** | C96-1058 | Three New Probabilistic Models For Dependency Parsing ... | 1996 | 66 |
| | P97-1003 | Three Generative, Lexicalized Models For Statistical Parsing | 1997 | 50 |
| | P99-1065 | A Statistical Parser For Czech | 1999 | 54 |
| | P05-1013 | Pseudo-Projective Dependency Parsing | 2005 | 40 |
| | P05-1012 | On-line Large-Margin Training Of Dependency Parsers | 2005 | 71 |
| **PBMT** | N03-1017 | Statistical Phrase-Based Translation | 2003 | 172 |
| | W03-0301 | An Evaluation Exercise For Word Alignment | 2003 | 11 |
| | J04-4002 | The Alignment Template Approach To Statistical Machine Translation | 2004 | 49 |
| | N04-1033 | Improvements In Phrase-Based Statistical Machine Translation | 2004 | 23 |
| | P05-1033 | A Hierarchical Phrase-Based Model For Statistical Machine Translation | 2005 | 65 |
| **Summ** | A00-1043 | Sentence Reduction For Automatic Text Summarization | 2000 | 19 |
| | A00-2024 | Cut And Paste Based Text Summarization | 2000 | 20 |
| | C00-1072 | The Automated Acquisition Of Topic Signatures ... | 2000 | 19 |
| | W00-0403 | Centroid-Based Summarization Of Multiple Documents ... | 2000 | 28 |
| | W03-0510 | The Potential And Limitations Of Automatic Sentence Extraction ... | 2003 | 14 |
| **QA** | A00-1023 | A Question Answering System Supported By Information Extraction | 2000 | 13 |
| | W00-0603 | A Rule-Based Question Answering System For Reading ... | 2002 | 19 |
| | P02-1006 | Learning Surface Text Patterns For A Question Answering System | 2002 | 72 |
| | D03-1017 | Towards Answering Opinion Questions: Separating Facts From Opinions ... | 2003 | 39 |
| | P03-1001 | Offline Strategies For Online Question Answering ... | 2003 | 27 |
| **TE** | D04-9907 | Scaling Web-Based Acquisition Of Entailment Relations | 2004 | 12 |
| | H05-1047 | A Semantic Approach To Recognizing Textual Entailment | 2005 | 7 |
| | H05-1079 | Recognising Textual Entailment With Logical Inference | 2005 | 9 |
| | W05-1203 | Measuring The Semantic Similarity Of Texts | 2005 | 17 |
| | P05-1014 | The Distributional Inclusion Hypotheses And Lexical Entailment | 2005 | 10 |
| **CRF** | N03-1023 | Weakly Supervised Natural Language Learning ... | 2003 | 29 |
| | N04-1042 | Accurate Information Extraction from Research Papers ... | 2004 | 24 |
| | W05-0622 | Semantic Role Labelling with Tree CRFs | 2005 | 9 |
| | P06-1009 | Discriminative Word Alignment with Conditional Random Fields | 2006 | 33 |
| | W06-1655 | A Hybrid Markov/Semi-Markov CRF for Sentence Segmentation | 2006 | 20 |

Table 1: Papers were extracted from 6 different NLP topics in AAN: Dependency Parsing (DP), Phrase-based Machine Translation (PBMT), Text Summarization (Summ), Question Answering (QA), Textual Entailment (TE), and Conditional Random Fields (CRF). Each set consists of the 5 highest cited papers in AAN's 2008 release whose title and content matched the corresponding topic phrase.

Below we describe our approach to citation analysis, including our calculation of inter-judge agreement. We then describe our C-LexRank method for extracting citation sentences.

## 3.1 Citation Analysis

To analyze citations, we designed an annotation task that requires explicit definitions that distinguish between phrases that represent the same or different information units. Unfortunately, there is little consensus in the literature on such definitions. Therefore, following van Halteren and Teufel (2003), Qazvinian and Radev (2011b) we made the following distinction. We define a *nugget* to be a phrasal information unit (i.e., any phrase that would contain some information about the contributions of the cited paper). Different nuggets may all represent the same atomic semantic unit, which we refer to as a *factoid*. In the context of citations, a factoid refers to a unique contribution of a target paper mentioned in a citation sentence. For example, the following set of citations to Eisner's (1996) famous parsing paper illustrate the set of factoids about this paper and suggest that different authors who cite a particular paper may discuss different contributions (factoids) of that paper.

*In the context of DPs, this <u>edge based factorization</u> method was proposed by* **Eisner (1996)**.

**Eisner (1996)** *gave a <u>generative model</u> with a <u>cubic parsing</u> algorithm based on an edge factorization of trees.*

**Eisner (1996)** *proposed an <u>$O(n^3)$ parsing algorithm for PDG</u>.*

*If the parse has to be projective,* **Eisner's (1996)** <u>*bottom-up-span algorithm*</u> *can be used for the search.*

This example also suggests that different authors use different wordings (nuggets) to represent the same factoids. For instance, *cubic parsing* and $O(n^3)$ *parsing algorithm* are two nuggets that represent the same factoid about (Eisner, 1996). A similar example, which we will use throughout the paper, is the paper by Cohn and Blunsom (2005) (identified with the ACL ID W05-0622 in Table 1). This paper is cited in 9 different sentences within AAN. All of these sentences are listed in Table 2. In each sentence, the nuggets extracted by the annotators are underlined. As this table suggests, a citation sentence may not discuss any of the contributions of the cited paper. For instance, the last sentence does not contain any factoids about Cohn and Blunsom's (2005) work. The nuggets that are identified using the citation to the paper (Cohn & Blunsom, 2005) account for a total number of 3 factoids (contributions) identified for this paper: $f_1$, tree structures; $f_2$, semantic role labeling; and $f_3$, a pipelined approach.

Following these examples, we asked two annotators with background in Natural Language Processing to review each citing sentence and extract a list of phrases that represent a contribution of the cited paper.[6] Moreover, to ensure that the extracted nuggets are explicitly mentioned in the citations, we asked the annotators to rely merely on the set of citations to do the task and not on their background on this topic or the source of the cited paper. Finally, we reviewed each list and collapsed phrases that represent the same contribution (factoid).

---

6. One of the annotators is an author of this paper.

Finding agreement between annotated well-defined nuggets is straightforward and can be calculated in terms of Kappa. However, when nuggets themselves are to be extracted by annotators, the task becomes less obvious. To calculate the agreement, we annotated 5 randomly selected citation sets twice (1 paper from each of the NLP areas in Table 1), and designed a simple evaluation scheme based on Kappa. For each $n$-gram, $w$, in a given citation sentence, we determine $w$ is part of any nugget in either human annotations. If $w$ occurs in both or neither, then the two annotators agree on it, and otherwise they do not. Based on this agreement setup, we can formalize the $\kappa$ statistic as:

$$\kappa = \frac{\Pr(a) - \Pr(e)}{1 - \Pr(e)} \tag{1}$$

where $Pr(a)$ is the relative observed agreement among annotators, and $Pr(e)$ is the probability that annotators agree by chance if each annotator is randomly assigning categories.

Table 3 shows the unigram, bigram, and trigram-based $\kappa$ between the two human annotators (**Human1, Human2**) in the five datasets that were annotated twice. These results suggest that human annotators can reach substantial agreement when trigram nuggets are examined, and have reasonable agreement for unigram and bigram nuggets.

## 3.2 C-LexRank

In this section we describe C-LexRank as a method to extract citing sentences that cover a diverse set of factoids. Our method works by modeling the set of citations as a network of sentences and identifying communities of sentences that cover similar factoids. Once a good division of sentences is made, we extract salient sentences from different communities. Figure 1 illustrates a representative example that depicts C-LexRank's process.

### 3.2.1 CITATION SUMMARY NETWORK

In the first step (as shown in Figure 1 (a)), we model the set of sentences that cite a specific paper with a network in which vertices represent citing sentences and undirected weighted edges show the degree of semantic relatedness between vertex pairs, normally quantified by a similarity measure. We refer to this network as the *Citation Summary Network* of an article. The similarity function should ideally assign high scores to sentence pairs that have the same factoids, and should assign low scores to sentences that talk about different contributions of the target paper.

Previously, Qazvinian and Radev (2008) examined 7 different similarity measures including TF-IDF with various IDF databases, longest common sub-sequence, generation probability (Erkan, 2006), and the Levenstein distance on a training set of citations. They showed that the cosine similarity measure that employs TF-IDF vectors assigns higher similarities to pairs that contain the same factoids. Following Qazvinian and Radev (2008), we use the cosine similarity between TF-IDF vector models that employ a general IDF corpus[7] to construct the citation summary network of each article.

---

7. We use the IDF corpus in the Mead summarization system (Radev et al., 2004), which is generated using the English portion of the Hong Kong News parallel corpus (Ma, 2000).

|   | factoid | Citation Sentence |
|---|---------|-------------------|
| 1 | $f_1$ | Our parsing model is based on a conditional random field model, however, unlike previous <u>TreeCRF</u> work, e.g., (**Cohn & Blunsom, 2005**; Jousse et al., 2006), we do not assume a particular tree structure, and instead find the most likely structure and labeling. |
| 2 | $f_3$ | Some researchers (Xue & Palmer, 2004; Koomen et al., 2005; **Cohn & Blunsom, 2005**; Punyakanok et al., 2008; Toutanova et al., 2005, 2008) used a <u>pipelined approach</u> to attack the task. |
| 3 | $f_2$ | They have been used for tree labelling, in XML tree labelling (Jousse et al., 2006) and <u>semantic role labelling</u> tasks (**Cohn & Blunsom, 2005**). |
| 4 | $f_1$ | Finally, probabilistic models have also been applied to produce the structured output, for example, generative models (Thompson, Levy, & Manning, 2003), sequence tagging with classifiers (Màrquez et al., 2005; Pradhan et al., 2005), and Conditional Random Fields on <u>tree structures</u> (**Cohn & Blunsom 2005**). |
| 5 | $f_3$ | As for SRL on news, most researchers used the <u>pipelined approach</u>, i.e., dividing the task into several phases such as argument identification, argument classification, global inference, etc., and conquering them individually (Xue & Palmer, 2004; Koomen et al., 2005; **Cohn & Blunsom**, 2005; Punyakanok et al., 2008; Toutanova et al., 2005, 2008). |
| 6 | $f_1, f_2$ | Although <u>T-CRF</u>s are relatively new models, they have already been applied to several NLP tasks, such as <u>semantic role labeling</u>, semantic annotation, word sense disambiguation, image modeling (**Cohn & Blunsom, 2005**; Tang et al., 2006; Jun et al., 2009; Awasthi et al., 2007). |
| 7 | $f_2$ | The model can be used for tasks like syntactic parsing (Finkel et al., 2008) and <u>semantic role labeling</u> (**Cohn & Blunsom, 2005**). |
| 8 | $f_1$ | Regarding novel learning paradigms not applied in previous shared tasks, we find Relevant Vector Machine (RVM), which is a kernel-based linear discriminant inside the framework of Sparse Bayesian Learning (Johansson & Nugues, 2005) and Tree Conditional Random Fields (<u>T-CRF</u>) (**Cohn & Blunsom, 2005**), that extend the sequential CRF model to tree structures. |
| 9 | N/A | We use CRFs as our models for both tasks (**Cohn & Blunsom, 2005**). |

Table 2: The AAN paper W05-0622 on CRF by Cohn & Blunsom (2005) is cited in nine different AAN sentences. In each citation sentence, the nuggets extracted by the annotators are underlined.

|  | **Average $\kappa$** | | |
|---|---|---|---|
|  | unigram | bigram | trigram |
| **Human1 vs. Human2** | | | |
| A00-1023 | 1.000 | 0.615 | 0.923 |
| H05-1079 | 0.889 | 0.667 | 0.556 |
| P05-1013 | 0.427 | 0.825 | 0.975 |
| W03-0301 | 0.455 | 0.636 | 0.818 |
| W05-0622 | 0.778 | 0.667 | 0.778 |
| **Average** | 0.710 | 0.682 | 0.810 |

Table 3: Agreement between different annotators in terms of Kappa in 5 citation sets.

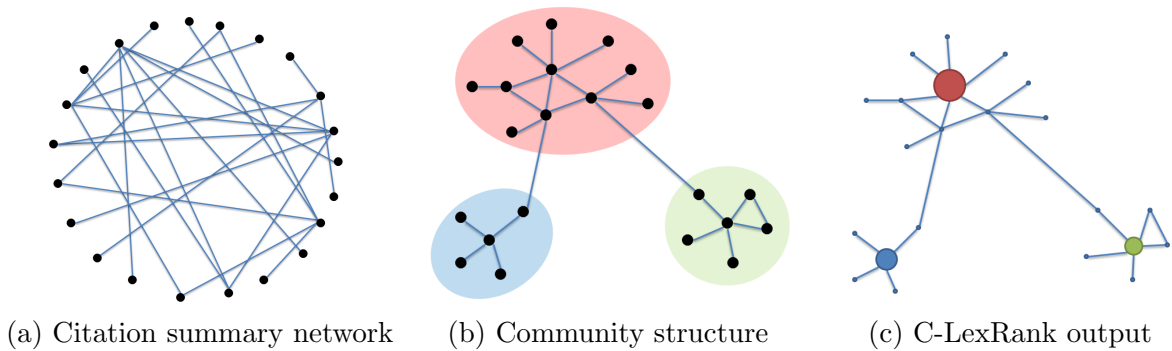

(a) Citation summary network     (b) Community structure     (c) C-LexRank output

Figure 1: The C-LexRank method extracts citing sentences that cover a diverse set of factoids. The citation summary network in (a) models the set of sentences that cite a specific paper, where vertices represent citing sentences and (weighted) edges show the degree of semantic relatedness between vertex pairs. The community structure in (b) corresponds to clustered sets of representative sentences extracted from citation sentences. The C-LexRank output in (c) corresponds to the candidate sentences from different clusters that are used for building a summary.

### 3.2.2 COMMUNITY STRUCTURE

In the second step (as shown in Figure 1 (b)), we extract vertex communities from the citation summary network to generate summaries. We generate summaries by extracting representative sentences from the citation summary network. Intuitively, a good summary should include sentences that represent different contributions of a paper. Therefore, a good sentence selection from the citation summary network will include vertices that are similar to many other vertices and which are not very similar to each other. On the other hand, a bad selection would include sentences that are only representing a small set of vertices in the graph. This is very similar to the concept of maximizing social influence in social networks (Kempe, Kleinberg, & Éva Tardos, 2003). Figure 2 shows an example in which the selected two vertices in the citation summary networks represent a small subset of vertices (left) and a larger subset of vertices (right). In our work we try to select vertices that maximize the size of the set of vertices that they represent. We achieve this by detecting different vertex communities in the citation summary network.

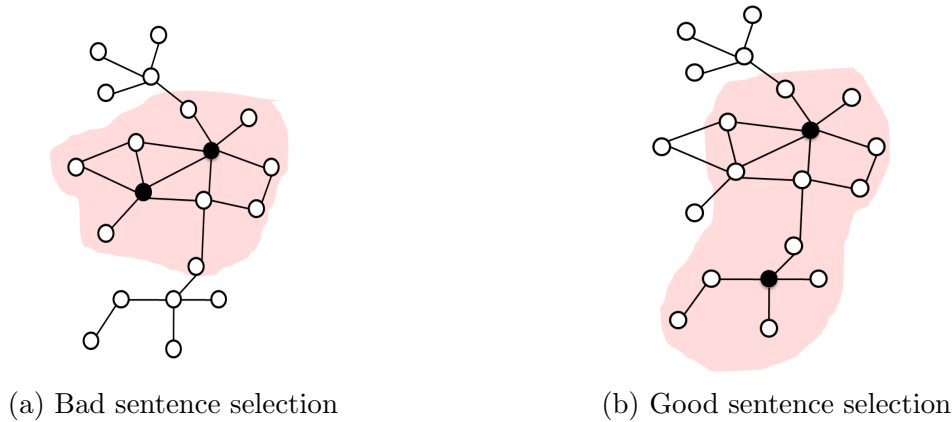

(a) Bad sentence selection          (b) Good sentence selection

Figure 2: Summaries are produced by using vertex coverage to select a set of representative vertices corresponding to sentences. Selecting two similar vertices will cause the summary to cover fewer contributions of the target paper in (a), while selecting less similar vertices as the summary will increase the coverage of the summary (b).

In order to find vertex communities and thus a good sentence selection, we exploit the small-world property of citation summary networks. A network is called *small-world*, if most of its vertices are not neighbors of each other, but can be reached from one another by a small number of steps (Watts & Strogatz, 1998). Recent research has shown that a wide range of natural graphs such as biological networks (Ravasz, Somera, Mongru, Oltvai, & Barabási, 2002), food webs (Montoya & Solé, 2002), brain neurons (Bassett & Bullmore, 2006) and human languages (Ferrer i Cancho & Solé, 2001) exhibit the small-world property.

This common characteristic can be detected using two basic statistical properties: the clustering coefficient $C$, and the average shortest path length $\ell$. The clustering coefficient of a graph measures the number of closed triangles in the graph. It describes how likely it is that two neighbors of a vertex are connected (Newman, 2003). Watts and Strogatz (1998) define the clustering coefficient as the average of the local clustering values for each vertex.

$$C = \frac{\sum_{i=1}^{n} c_i}{n} \qquad (2)$$

The local clustering coefficient $c_i$ for the $i$th vertex is the number of triangles connected to vertex $i$ divided by the total possible number of triangles connected to vertex $i$. Watts and Strogatz (1998) show that small-world networks are highly clustered and obtain relatively short paths (i.e., $\ell$ is small). Previous work (Qazvinian & Radev, 2011a) shows that citation summary networks are highly clustered. These networks have small shortest paths and obtain clustering coefficient values that are significantly larger than random networks. Moreover, Qazvinian and Radev suggest that this is because of a community structure, where each community is composed of a set of highly connected vertices with a small number of edges that fall between communities.

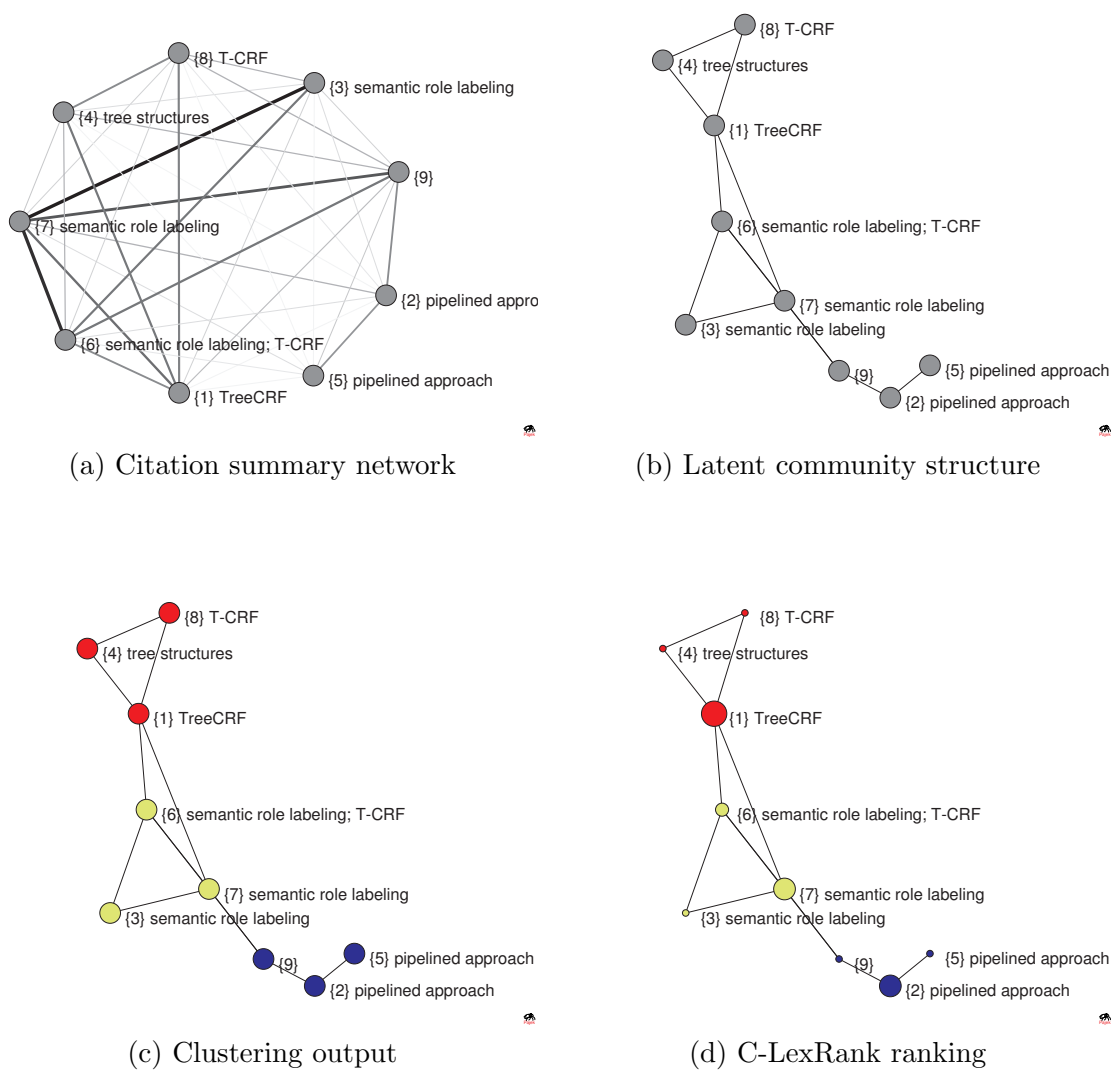

(a) Citation summary network

(b) Latent community structure

(c) Clustering output

(d) C-LexRank ranking

Figure 3: The C-LexRank algorithm operates as follows on Cohn and Blunsom's (2005) citation summary network: In the network (a), the vertices are citation sentences (annotated with their nuggets from Table 2), and each edge is the cosine similarity between the corresponding node pairs. (b) shows that the network has an underlying structure which is captured by C-LexRank in (c). Finally, (d) shows the C-LexRank output where vertex size is proportional to its LexRank value within the cluster.

Figure 3 (a) illustrates a real citation summary network built using the citation sentences in Table 2 in which each vertex is labeled with its corresponding nugget. With some re-arrangement of the vertices in Figure 3 (b), it becomes clear that the citation summary network of this paper has an underlying community structure in which sentences that cover similar factoids are closer to each other and form communities. For instance, in this network there are at least 3 observable communities: one that is about "$f_1$: tree structure," one about "$f_2$: semantic role labeling" and the last one about the "$f_3$: pipelined approach" as proposed by Cohn and Blunsom (2005).

In order to detect these communities automatically we use modularity. *Modularity*, (Newman, 2004a), is a measure to evaluate the divisions that a community detection algorithm generates. For a division with $g$ groups, they define matrix $\mathbf{e}_{g \times g}$ whose component $e_{ij}$ is the fraction of edges in the original network that connect vertices in components $i$, $j$. Then the modularity $Q$ can be defined as:

$$Q = \sum_i e_{ii} - \sum_{ijk} e_{ij} e_{ki} \tag{3}$$

Intuitively, $Q$ is the fraction of all the edges that are embedded within communities minus the expected value of the same quantity in a network with the same degrees but in which edges are placed at random regardless of the community structure. Newman and Girvan (2004) and Newman (2004b) showed across a wide range of simulated and real-world networks that larger $Q$ values are correlated with better graph clusterings. It is also shown by Newman (2004b) that if no edges exist that connect vertices across different clusters then $Q = 1$, and conversely if the number of inter-cluster edges is no better than random then $Q = 0$. Other work (Smyth & White, 2005) showed empirically that modularity works well in practice in terms of both (a) finding good clusterings of vertices in networks where community structure is evident, and (b) indicating what the appropriate number of clusters $k$ is for such a graph.

C-LexRank uses the clustering algorithm of Clauset, Newman, and Moore (2004), which exploits modularity to detect vertex communities in a network. This network clustering method, as discussed by Clauset et al. (2004) is a hierarchical agglomeration algorithm, which works by greedily optimizing the modularity in a linear running time for sparse graphs. More particularly, their method continuously merges vertex or cluster pairs with the highest similarity and stops when modularity reaches the maximum value. This clustering algorithm is efficient ($O(n \log^2 n)$ in the number of nodes, $n$) and does not require a predetermined number of clusters. These two characteristics makes this community detection algorithm particularly useful.

Figure 3 (c) shows how the clustering algorithm detects factoid communities in Cohn and Blunsom's (2005) citation summary network. In this figure, we have color-coded vertices based on their community. The clustering algorithm assigns sentences 1, 4 and 8 (which are all about the tree structures) to one cluster; sentences 3, 6 and 7 (which are all about semantic role labeling) to another cluster; and finally assigns sentences 2, 5 and 9 (sentences 2 and 5 are both about pipelined approach) to the last cluster. This figure also shows that sentence 6, which discusses two factoids ("semantic role labeling" and "T-CRF") connects the two vertex communities (corresponding to 2 factoids) as a bridge.

To evaluate how well the clustering method works in all of our datasets, we calculated both the *purity* and the *normalized mutual information* (NMI) for the divisions in each citation set, extracted using the community detection algorithm. Purity (Zhao & Karypis, 2001) is a method in which each cluster is assigned to the class with the majority vote in the cluster, and the accuracy of this assignment is then measured by dividing the number of correctly assigned documents by $N$. More formally:

$$\text{purity}(\Omega, \mathbb{C}) = \frac{1}{N} \sum_k \max_j |\omega_k \cap c_j| \tag{4}$$

where $\Omega = \{\omega_1, \omega_2, \ldots, \omega_K\}$ is the set of clusters and $\mathbb{C} = \{c_1, c_2, \ldots, c_J\}$ is the set of classes. $\omega_k$ is interpreted as the set of documents in the cluster $\omega_k$ and $c_j$ as the set of documents in the class $c_j$.

We also calculate the *normalized mutual information* (NMI). Manning, Raghavan, and Schütze (2008) describe NMI as follows. Let us assume $\Omega = \{\omega_1, \omega_2, \ldots, \omega_K\}$ is the set of clusters and $\mathbb{C} = \{c_1, c_2, \ldots, c_J\}$ is the set of classes. Then,

$$\text{NMI}(\Omega, \mathbb{C}) = \frac{I(\Omega; \mathbb{C})}{[H(\Omega) + H(\mathbb{C})]/2} \tag{5}$$

where $I(\Omega; \mathbb{C})$ is the mutual information:

$$I(\Omega, \mathbb{C}) = \sum_k \sum_j P(\omega_k \cap c_j) \log \frac{P(\omega_k \cap c_j)}{P(\omega_k)P(c_j)} \tag{6}$$

$$= \sum_k \sum_j \frac{|\omega_k \cap c_j|}{N} \log \frac{N|\omega_k \cap c_j|}{|\omega_k||c_j|} \tag{7}$$

in which $P(\omega_k)$, $P(c_j)$, and $P(\omega_k \cap c_j)$ are the probabilities of a document being in cluster $\omega_k$, class $c_j$, and in the intersection of $\omega_k$ and $c_j$, respectively; and $H$ is entropy:

$$H(\Omega) = -\sum_k P(\omega_k) \log P(\omega_k) \tag{8}$$

$$= -\sum_k \frac{|\omega_k|}{N} \log \frac{|\omega_k|}{N} \tag{9}$$

$I(\Omega; \mathbb{C})$ in Equation 6 measures the amount of information that we would lose about the classes without the cluster assignments. The normalization factor $([H(\Omega) + H(\mathbb{C})]/2)$ in Equation 5 enables us to trade off the quality of the clustering against the number of clusters, since entropy tends to increase with the number of clusters. For example, $H(\Omega)$ reaches its maximum when each document is assigned to a separate cluster. Because NMI is normalized, we can use it to compare cluster assignments with different numbers of clusters. Moreover, $[H(\Omega) + H(\mathbb{C})]/2$ is a tight upper bound for $I(\Omega; \mathbb{C})$, making NMI obtain values between 0 and 1. Table 4 lists the average Purity and NMI across the papers in our collected dataset, along with the analogous numbers for a division of the same size where vertices are randomly assigned to clusters.

### 3.2.3 RANKING

The third step of the C-LexRank process (as shown in Figure 1 (c)) is applied after the graph is clustered and the communities are formed. To produce the C-LexRank output, we extract sentences from different clusters to build a summary. We start with the largest cluster and extract sentences using LexRank (Erkan & Radev, 2004) within each cluster. In other words, for each cluster $\Omega_i$ we made a lexical network of *the sentences in that cluster* ($N_i$). Using LexRank we can find the most central sentences in $N_i$ as salient sentences of $\Omega_i$ to include in the main summary. We choose, for each cluster $\Omega_i$, the most salient sentence of $\Omega_i$, and if we have not reached the summary length limit, we do that for the second most salient sentences of each cluster, and so on. The cluster selection is in order of decreasing size. Figure 3 (d) shows Cohn and Blunsom's (2005) citation summary network, in which each vertex is plotted with a size proportional to its LexRank value within its cluster. This figure shows how C-LexRank emphasizes on selecting a diverse set of sentences covering a diverse set of factoids.

|  | average | 95% Confidence Interval |
|---|---|---|
| purity$(\Omega, \mathbb{C})$ | **0.461** | [0.398, 0.524] |
| purity$(\Omega_{\text{random}}, \mathbb{C})$ | 0.389 | [0.334, 0.445] |
| NMI$(\Omega, \mathbb{C})$ | **0.312** | [0.251, 0.373] |
| NMI$(\Omega_{\text{random}}, \mathbb{C})$ | 0.182 | [0.143, 0.221] |

Table 4: The average purity (in boldface) and normalized mutual information (NMI) values are shown for the papers in our collected dataset, along with analogous values for a division of the same size where vertices are randomly assigned to clusters.

Previously, we mentioned that factoids with higher weights appear in a greater number of sentences, and clustering aims to cluster such fact-sharing sentences in the same communities. Thus, starting with the largest community is important to ensure that the system summary first covers the factoids that are more frequently mentioned in other citation sentences and thus are more important.

The last sentence in the example in Table 2 is as follows. "We use CRFs as our models for both tasks (Cohn & Blunsom, 2005)." This sentence shows that a citation may not cover any contributions of the target paper. Such sentences are assigned by the community detection algorithm in C-LexRank to clusters to which they are semantically most similar. The intuition behind employing LexRank within each cluster is to try to avoid extracting such sentences for the summary, since LexRank within a cluster enforces extracting the most central sentence in that cluster. In order to verify this, we also try a variant of C-LexRank in which we do not select sentences from clusters based on their salience in the cluster, but rather in a round-robin fashion, in which all the sentences within a cluster are equally likely to be selected. We call this variant *C-RR*.

Table 5 shows the 100-word summary constructed using C-LexRank for our exemplar paper, in which different nuggets are illustrated in bold. This summary is a perfect summary in terms of covering the different factoids about the paper. It includes citing sentences that talk about "tree CRF," "pipelined approach," and "Semantic Role Labeling," which are indeed Cohn and Blunsom's (2005) three main contributions.

Our parsing model is based on a conditional random field model, however, unlike previous **TreeCRF** work, e.g., (Cohn & Blunsom, 2005; Jousse et al., 2006), we do not assume a particular **tree structure**, and instead find the most likely structure and labeling.

Some researchers (Xue & Palmer, 2004; Koomen et al., 2005; Cohn & Blunsom, 2005; Punyakanok et al., 2008; Toutanova et al., 2005, 2008) used a **pipelined approach** to attack the task.

The model can be used for tasks like syntactic parsing (Finkel et al., 2008) and **Semantic Role Labeling** (Cohn & Blunsom, 2005).

Table 5: This 100-word summary was constructed using C-LexRank for Cohn and Blunsom's (2005) citation summary network. Factoids are shown in bold face.

## 4. Other Methods

In our experiments in Section 5 we compare C-LexRank to a number of other summarization systems. We compare C-LexRank with random summarization to find a lower-bound on the pyramid scores in our experiments. We use LexRank and C-RR as two variants of C-LexRank to investigate the usefulness of community detection and salient vertex selection in C-LexRank. We evaluate DivRank as a state of the art graph-based summarization system that leverages diversity as well as MMR as a widely used diversity-based summarization system. Finally, we use Multiple Alternate Sentence Compression Summarizer (MASCS) as a system that does not rely merely on extraction, but rather produces a list of candidates by applying pre-defined sentence-compression rules.

### 4.1 Random

This method simply chooses citations in random order without replacement. Since a citing sentence may cover no information about the cited paper (as in the last sentence in Table 2), randomization has the drawback of selecting citations that have no valuable information in them. Moreover, the random selection procedure is more prone to produce redundant summaries as citing sentences that discuss the same factoid may be selected.

### 4.2 LexRank

One of the systems that we compare C-LexRank with is LexRank (Erkan & Radev, 2004). It works by first building a graph of all the documents ($d_i$) in a cluster. The edges between corresponding vertices ($d_i$) represent the cosine similarity between them if the cosine value is above a threshold (0.10 following Erkan & Radev, 2004). Once the network is built, the system finds the most central sentences by performing a random walk on the graph.

$$p(d_j) = (1 - \lambda)\frac{1}{|D|} + \lambda \sum_{d_i} p(d_i) P(d_i \to d_j) \tag{10}$$

Equation 10 shows that the probability that the random walker would visit $d_j$ depends on a random jump element as well as the probability that the random walk visits each of its neighbors ($d_i$) times the transition probability between $d_i$ and $d_j$, $P(d_i \rightarrow d_j)$.

Comparing C-LexRank summaries with the ones from LexRank gives insight into how we can benefit from detecting communities in citation sets. Essentially, C-LexRank is the same as LexRank if all the vertices are assigned to the same cluster. By construction, C-LexRank should produce more diverse summaries covering different perspectives by capturing communities of sentences that discuss the same factoids.

### 4.3 MMR

Maximal Marginal Relevance (MMR) is proposed by Carbonell and Goldstein (1998) and is a widely used algorithm in generating summaries that reflect the diversity of perspectives in the source documents (Das & Martins, 2007). MMR uses the pairwise cosine similarity matrix and greedily chooses sentences that are the least similar to those already in the summary. In particular,

$$MMR = \mathrm{argmin}_{D_i \in D - A} \left[ \max_{D_j \in A} Sim(D_i, D_j) \right] \qquad (11)$$

where $A$ is the set of documents in the summary, initialized to $A = \emptyset$. In Equation 11, a sentence $D_i$ that is not in the summary $A$ is chosen such that its highest similarity to summary sentences $\max_{D_j \in A} Sim(D_i, D_j)$ is minimum among all unselected sentences.

### 4.4 DivRank

We also compare C-LexRank with a state-of-the-art graph-based summarization system that leverages diversity, DivRank. DivRank is based on calculating stationary distribution of vertices using a modified random walk model. Unlike other time-homogeneous random walks (e.g., PageRank), DivRank does not assume that the transition probabilities remain constant over time.

DivRank uses a *vertex-reinforced random walk* model to rank graph vertices based on a diversity based centrality. The basic assumption in DivRank is that the transition probability from a vertex to other is reinforced by the number of previous visits to the target vertex (Mei et al., 2010). Particularly, let us assume $p_T(u, v)$ is the transition probability from any vertex $u$ to vertex $v$ at time $T$. Then,

$$p_T(d_i, d_j) = (1 - \lambda).p^*(d_j) + \lambda.\frac{p_0(d_i, d_j).N_T(d_j)}{D_T(d_i)} \qquad (12)$$

where $N_T(d_j)$ is the number of times the walk has visited $d_j$ up to time $T$ and,

$$D_T(d_i) = \sum_{d_j \in V} p_0(d_i, d_j) N_T(d_j) \qquad (13)$$

Here, $p^*(d_j)$ is the prior distribution that determines the preference of visiting vertex $d_j$, and $p_0(u, v)$ is the transition probability from $u$ to $v$ prior to any reinforcement. Mei et al. argue that the random walk could stay at the current state at each time, and therefore assumes a hidden link from each vertex to itself, thus defining $p_0(u, v)$ as:

$$p_0(u, v) = \begin{cases} \alpha . \frac{w(u,v)}{deg(u)} & \text{if } u \neq v \\ 1 - \alpha & \text{if } u = v \end{cases} \tag{14}$$

Here, we try two variants of this algorithm: *DivRank*, in which $p^*(d_j)$ is uniform, and *DivRank with priors* in which $p^*(d_j) \propto l(D_j)^{-\beta}$, where $l(D_j)$ is the number of the words in the document $D_j$ and $\beta$ is a parameter (0.1 in our experiments). This prior distribution assigns larger probabilities to shorter sentences which will increase their likelihood of being salient. This will cause more sentences to be included in the summary, and might increase the factoid coverage. In our experiments, we follow Mei et al. (2010) and set $\lambda = 0.90$ and $\alpha = 0.25$.

## 4.5 MASCS

The last summarization system that we use as a baseline is Multiple Alternate Sentence Compression Summarizer (MASCS) (Zajic, Dorr, Lin, & Schwartz, 2007). Similar to previous previous baseline systems, MASCS's goal is to leverage diversity in summarization. MASCS performs preprocessing on sentences that transforms them into new sentences, thus expanding the pool of candidates available for inclusion in a summary beyond the set of sentences that occur in source documents. This is what makes MASCS somewhat non-extractive. In addition, the preprocessing used in MASCS for these experiments was specifically adapted to the genre of citation sentences from scientific papers (Whidby, 2012).

More particularly, MASCS is a summarization system that utilizes Trimmer's (Zajic et al., 2007) sentence compression candidates to create summaries for a single or set of documents. Summarization with MASCS is performed in three stages. In the first stage, MASCS generates several compressed sentence candidates for every sentence in a document from the cluster. The second stage involves calculating various ranking features for each of the compressed sentence candidates. In the final stage, sentence candidates are chosen for inclusion in the summary, and are chosen based on a linear combination of features.

Trimmer can leverage the output of any constituency parser that uses the Penn Treebank conventions. At present, the Stanford Parser (Klein & Manning, 2003) is used. The set of compressions is ranked according to a set of features that may include metadata about the source sentences, details of the compression process that generated the compression, and externally calculated features of the compression.

Summaries are constructed by iteratively adding compressed sentences from the candidate pool until a length threshold is met. Candidates are chosen by an implementation of MMR that uses features directly calculated from a candidate, metadata about a candidate's source and its compression history, relevance of the candidate to the topic and redundancy of the candidate to already selected candidates. The redundancy score in MASCS uses an index of the words $(w)$ in the document set:

$$\sum_w \log(\lambda . P(w|\text{summary}) + (1 - \lambda) . P(w|\text{corpus})) \tag{15}$$

where $\lambda$ is a weighting factor (set to 0.3 in our experiments).

## 5. Experiments

We used the 30 sets of citations listed in Table 1 and employ C-LexRank to produce two extractive summaries with different summary lengths (100 and 200 words) for each set. In addition to C-LexRank and C-RR, we also performed the same experiments with the baseline methods described in Section 4, most of which are aimed at leveraging diversity in summarization.

### 5.1 Evaluation

To evaluate our system, we use the pyramid evaluation method (Nenkova & Passonneau, 2004). Each factoid in the citations to a paper corresponds to a *summary content unit (SCU)* in (Nenkova & Passonneau, 2004).

The score given by the pyramid method for a summary is the ratio of the sum of the weights of its factoids to the sum of the weights of an optimal summary. This score ranges from 0 to 1, and high scores show the summary content contains more heavily weighted factoids. If a factoid appears in more citation sentences than another factoid, it is more important, and thus should be assigned a higher weight. To weight the factoids we build a pyramid, and each factoid falls in a tier. Each tier shows the number of sentences a factoid appears in. Thus, the number of tiers in the pyramid is equal to the citation summary size. If a factoid appears in more sentences, it falls in a higher tier. So, if the factoid $f_i$ appears $|f_i|$ times in the citation summary it is assigned to the tier $T_{|f_i|}$.

The pyramid score formula that we use is computed as follows. Suppose the pyramid has $n$ tiers, $T_i$, where tier $T_n$ on top and $T_1$ on the bottom. The weight of the factoids in tier $T_i$ will be $i$ (i.e. they appeared in $i$ sentences). If $|T_i|$ denotes the number of factoids in tier $T_i$, and $D_i$ is the number of factoids in the *summary* that appear in $T_i$, then the total factoid weight for the summary is as follows.

$$D = \sum_{i=1}^{n} i \times D_i \tag{16}$$

Additionally, the optimal pyramid score for a summary with $X$ factoids, is

$$Max = \sum_{i=j+1}^{n} i \times |T_i| + j \times (X - \sum_{i=j+1}^{n} |T_i|) \tag{17}$$

where $j = \max_i(\sum_{t=i}^{n} |T_t| \geq X)$. Subsequently, the pyramid score for a summary is calculated as follows.

$$P = \frac{D}{Max} \tag{18}$$

### 5.2 Results and Discussion

Table 6 shows the average pyramid score of the summaries generated using different methods with different lengths. Longer summaries result in higher pyramid scores since the amount of information they cover is greater than shorter summaries. For the random sentence extraction baseline, we repeat the experiment with 100 different randomly generated seed values and report the average pyramid score of these summaries in Table 6. This table shows that C-LexRank outperforms all other methods that leverage diversity as well as random summaries and LexRank. The results in this table also suggest that employing LexRank within each cluster is essential for the selection of salient citing sentences, as the average pyramid scores from C-RR, where sentences are picked in a round-robin fashion, are lower.

| | Length: 100 words | | Length: 200 words | |
|---|---|---|---|---|
| Method | pyramid | 95% C.I. | pyramid | 95% C.I. |
| Random | 0.535 | [0.526,0.645] | 0.748 | [0.740,0.755] |
| MMR | 0.600 | [0.501,0.699] | 0.761 | [0.685,0.838] |
| LexRank | 0.604 | [0.511,0.696] | 0.784 | [0.725,0.844] |
| DivRank | 0.644 | [0.580,0.709] | 0.769 | [0.704,0.834] |
| DivRank (with priors) | 0.632 | [0.545,0.719] | 0.778 | [0.716,0.841] |
| MASCS | 0.571 | [0.477,0.665] | 0.772 | [0.705,0.840] |
| C-RR | 0.513 | [0.436,0.591] | 0.755 | [0.678,0.832] |
| C-LexRank | **0.647** | **[0.565,0.730]** | **0.799** | **[0.732,0.866]** |
| C.I.=Confidence Interval | | | | |

Table 6: Average pyramid scores are shown for two different summary lengths (100 words and 200 words) for eight different methods, including a summary generator based on random citation sentence selection. C-LexRank outperforms all other methods that leverage diversity as well as random summaries and LexRank. Highest scores for each input source are shown in bold.

### 5.2.1 EFFECT OF COMMUNITY DETECTION

The community detection that C-LexRank employs assigns highly similar citing sentences to the same cluster. This enables C-LexRank to produce a diverse summary by selecting sentences from different clusters. This selection is done by assigning a score to each vertex using LexRank within the cluster. The modularity-based clustering method described in Section 3.2.2, which works by maximizing modularity in a clustering, will always produce at least 2 clusters. Intuitively, in a network in which all the vertices are assigned to the same community, the fraction of edges that are embedded within that community is equal to the expected value of the same quantity in a network in which edges are placed at random. This will make $Q$ obtain its lower-bound, $Q = 0$.

However, in the hypothetical case where all the vertices belong to the same cluster, C-LexRank will be the same as LexRank (i.e., it will perform LexRank on the entire network). Therefore, comparing C-LexRank and LexRank helps us understand the effect of clustering on summary quality. Table 6 shows that C-LexRank produces summaries that obtain higher pyramid scores at both 100 words and 200 words. Table 7 shows the 100-word summary that was constructed using LexRank for Cohn and Blunsom's (2005) citations. This summary, unlike the one produced by C-LexRank (Table 5), does not cover all of the factoids of the target paper (e.g., "pipelined approach"). Moreover, this summary has redundant information (e.g., TreeCRF vs. T-CRF) and contains a citation sentence that does not cover any of the contributions of Cohn and Blunsom.

The model can be used for tasks like syntactic parsing (Finkel et al., 2008) and **semantic role labeling** (Cohn & Blunsom, 2005).
We use CRFs as our models for both tasks (Cohn & Blunsom, 2005).
Our parsing model is based on a conditional random field model, however, unlike previous **TreeCRF** work, e.g., (Cohn & Blunsom, 2005; Jousse et al., 2006), we do not assume a particular tree structure, and instead find the most likely structure and labeling.
Although **T-CRF**s are relatively new models, they have already been applied to several NLP tasks, such as **semantic role labeling**, semantic annotation, word sense disambiguation, image modeling (Cohn & Blunsom, 2005; Tang et al., 2006; Jun et al., 2009; Awasthi et al., 2007).

Table 7: The summary constructed using LexRank and Cohn and Blunsom's (2005) citation sentences. Compared to the C-LexRank summary (in Table 5), LexRank does not produce a summary of all Cohn and Blunsom's (2005) contributions (The summary is not truncated for clarity).

### 5.2.2 Salient Vertex Extraction

Selecting representative sentences (vertices) from different clusters is done using LexRank in the C-LexRank algorithm. More particularly, for each cluster, C-LexRank first extracts a subgraph of the network representing vertices and edges in that cluster, and then employs LexRank to assign a salience score to each vertex. An alternative idea could be selecting sentences from clusters at random (C-RR). In C-RR, we traverse between clusters in a round-robin fashion and randomly select a previously unselected sentence from the cluster to include a summary.

Comparing C-LexRank with C-RR enables us to understand the effect of salience selection within communities. Selecting vertices that are not a good representative of the cluster may result in picking sentences that do not cover contributions of the target paper (e.g., sentence 9 from Table 2 – vertex 9 in Figure 3 (d)). In fact, Table 6 shows that C-LexRank produces summaries with relatively 20% and 5% higher pyramid scores than C-RR when extracting 100 and 200 word summaries respectively. Moreover, C-RR performs better when longer summaries are produced since it extracts a greater number of sentences from each cluster increasing the likelihood of covering different factoids captured by different clusters.

## 6. Summaries of Scientific Topics

In previous sections, we described C-LexRank as a method to identify communities of citations that discuss the same scientific contributions. We showed that C-LexRank is effective in summarizing contributions of single scientific papers. However, the ultimate goal of our work is to investigate whether citations have summary-amenable information and also to build an end-to-end system that receives a query representing a scientific topic (such as "dependency parsing") and produces a citation-based automatic summary of the given topic.

In this section, we extend our experiments on using the tools explained in previous sections for automatic summarization of scientific topics. Our evaluation experiments for extractive summary generation are on a set of papers in the research area of Question Answering (QA) and another set of papers on Dependency Parsing (DP). The two sets of papers were compiled by selecting all the papers in AAN that had the words "Question Answering" and "Dependency Parsing," respectively, in the title and the content. There were 10 papers in the QA set and 16 papers in the DP set. We also compiled the citation sentences for the 10 QA papers and the citation sentences for the 16 DP papers.

## 6.1 Data Preparation

Our goal is to determine if citations do indeed have useful information that one will want to put in a summary and if so, how much of this information is *not* available in the original papers and their abstracts. For this we evaluate each of the automatically generated summaries using two separate approaches: nugget-based pyramid evaluation and ROUGE. Recall-Oriented Understudy for Gisting Evaluation (ROUGE) is a metric that evaluates automatic summaries by comparing them against a set of human written references (Lin, 2004).

Two sets of gold standard data were manually created from the QA and DP citation sentences and abstracts, respectively:[8] (1) We asked three annotators[9] with background in Natural Language Processing to identify important nuggets of information worth including in a summary. (2) We asked four NLP researchers[10] to write 250-word summaries of the QA and DP datasets. Then we determined how well the different automatically generated summaries perform against these gold standards. If the citations contain only redundant information with respect to the abstracts and original papers, then the summaries of citations will not perform better than others.

### 6.1.1 Nugget Annotations

For our first evaluation approach we used a nugget-based evaluation methodology (Voorhees, 2003; Nenkova & Passonneau, 2004; Hildebrandt, Katz, & Lin, 2004; Lin & Demner-Fushman, 2006). We asked three annotators with background in Natural Language Processing to review the citation sentences and/or abstract sets for each of the papers in the QA and DP sets and manually extract prioritized lists of 2–8 "factoids," or main contributions, supplied by each paper. Each factoid was assigned a weight based on the frequency with which it was listed by annotators as well as the priority it was assigned in each case. Our automatically generated summaries were then scored based on the number and weight of the nuggets that they covered.

More particularly, the annotators had two distinct tasks for the QA set, and one for the DP set: (1) extract nuggets for each of the 10 QA papers, based only on the citations to those papers; (2) extract nuggets for each of the 10 QA papers, based only on the abstracts of those papers; and (3) extract nuggets for each of the 16 DP papers, based only on the citations to those papers.

---

8. Creating gold standard data from complete papers is fairly arduous, and was not pursued.
9. Two of the annotators are authors of this paper.
10. All of the annotators are authors of this paper.

| | **Human Performance: Pyramid score** | | | | |
|---|---|---|---|---|---|
| | Human1 | Human2 | Human3 | Human4 | Average |
| **Input: QA citations** | | | | | |
| QA–CT nuggets | 0.524 | 0.711 | 0.468 | 0.695 | 0.599 |
| QA–AB nuggets | 0.495 | 0.606 | 0.423 | 0.608 | 0.533 |
| **Input: QA abstracts** | | | | | |
| QA–CT nuggets | 0.542 | 0.675 | 0.581 | 0.669 | 0.617 |
| QA–AB nuggets | 0.646 | 0.841 | 0.673 | 0.790 | 0.738 |
| **Input: DP citations** | | | | | |
| DP–CT nuggets | 0.245 | 0.475 | 0.378 | 0.555 | 0.413 |

Table 8: Pyramid scores were computed for human-created summaries of QA and DP data. The summaries were evaluated using nuggets drawn from QA citation sentences (QA–CT), QA abstracts (QA–AB), and DP citation sentences (DP–CT).

One annotator completed the three tasks in full and the remaining two annotators jointly completed tasks 1 and 3, providing us with two complete annotations of the QA and DP citation sets and one annotation of the QA abstract set. For each task, annotators constructed lists of 2–8 prioritized nuggets per paper. This gave us 81 distinct nuggets from the QA citation set, 45 nuggets from the QA abstract set, and 144 nuggets from the DP citation set. By collapsing similar nuggets, we were able to identify 34 factoids for the QA citation set, 27 factoids for the QA abstract set, and 57 factoids for the DP citation set. We obtained a weight for each factoid by reversing its priority out of 8 (e.g., a factoid listed with priority 1 was assigned a weight of 8, a nugget listed with priority 2 was assigned a weight of 7, etc.) and summing the weights over each listing of that factoid.[11]

### 6.1.2 Expert Summaries

In addition to nugget annotations, we asked four NLP researchers to write 250-word summaries of the QA citation set, QA abstract set and DP citation set.

Table 8 gives the pyramid scores of the 250-word summaries manually produced by experts. The summaries were evaluated using the nuggets drawn from the QA citations, QA abstracts, and DP citations. The average of their scores (listed in the rightmost column) may be considered a good score to aim for by the automatic summarization methods. Additionally, Table 9 presents ROUGE scores (Lin, 2004) of each of expert-written 250-word summaries against each other (e.g., Human1 versus all others and so forth). The average (last column) could be considered a ceiling in the performance of the automatic summarization systems.

---

11. Results obtained with other weighting schemes that ignored priority ratings and multiple mentions of a nugget by a single annotator showed the same trends as the ones shown by the selected weighting scheme.

| | **Human Performance: ROUGE-2** | | | | |
|---|---|---|---|---|---|
| | human1 | human2 | human3 | human4 | average |
| **Input: QA citations** | | | | | |
| QA–CT refs. | 0.181 | 0.196 | 0.076 | 0.202 | 0.163 |
| QA–AB refs. | 0.112 | 0.140 | 0.071 | 0.158 | 0.120 |
| **Input: QA abstracts** | | | | | |
| QA–CT refs. | 0.131 | 0.110 | 0.122 | 0.115 | 0.120 |
| QA-AB refs. | 0.265 | 0.198 | 0.180 | 0.254 | 0.224 |
| **Input: DP citations** | | | | | |
| DP–CT refs. | 0.155 | 0.126 | 0.120 | 0.165 | 0.142 |

Table 9: ROUGE-2 scores were obtained for each of the manually created summaries by using the other three as reference. ROUGE-1 and ROUGE-L followed similar patterns.

## 6.2 Automatic Extractive Summaries

We used four summarization systems for our summary-creation approach: *C-LexRank, C-RR, LexRank* and *MASCS*.

We automatically generated summaries for both QA and DP from three different types of documents: (1) full papers from the QA and DP sets—*QA and DP full papers (PA)*, (2) only the abstracts of the QA and DP papers—*QA and DP abstracts (AB)*, and (3) the citation sentences corresponding to the QA and DP papers—*QA and DP citations (CT)*.

We generated 24 ($4 \times 3 \times 2$) summaries, each of length 250 words, by applying MASCS, LexRank, and C-LexRank on the three data types (citation sentences, abstracts, and full papers) for both QA and DP. (Table 10 shows a fragment of one of the automatically generated summaries from QA citation sentences.) We created six ($3 \times 2$) additional 250-word summaries by randomly choosing sentences from citations, abstracts, and full papers of QA and DP. We will refer to them as *random summaries*.

Most of work in QA and paraphrasing focused on folding paraphrasing knowledge into question analyzer or answer locater (Rinaldi et al., 2003; Tomuro, 2003).

In addition, number of researchers have built systems to take reading comprehension examinations designed to evaluate children's reading levels (Charniak et al., 2000; Hirschman et al., 1999; Ng et al., 2000; Riloff & Thelen, 2000; Wang et al., 2000).

So-called " definition " or " other " questions at recent TREC evaluations (Voorhees, 2005) serve as good examples.

To better facilitate user information needs, recent trends in QA research have shifted towards complex, context-based, and interactive question answering (Voorhees, 2001; Small et al., 2003; Harabagiu et al., 2005).

Table 10: A fragment of one of the MASCS-generated summaries is illustrated here using the QA citation sentences as input.

| | **System Performance: Pyramid score** | | | | |
|---|---|---|---|---|---|
| | Random | C–LexRank | C–RR | LexRank | MASCS |
| **Input: QA citations** | | | | | |
| QA–CT nuggets | 0.321 | 0.434 | 0.268 | 0.295 | **0.616** |
| QA–AB nuggets | 0.305 | 0.388 | 0.349 | 0.320 | **0.543** |
| **Input: QA abstracts** | | | | | |
| QA–CT nuggets | 0.452 | 0.383 | **0.480** | 0.441 | 0.404 |
| QA–AB nuggets | **0.623** | 0.484 | 0.574 | 0.606 | 0.622 |
| **Input: QA full papers** | | | | | |
| QA–CT nuggets | 0.239 | **0.446** | 0.299 | 0.190 | 0.199 |
| QA–AB nuggets | 0.294 | **0.520** | 0.387 | 0.301 | 0.290 |
| **Input: DP citations** | | | | | |
| DP–CT nuggets | 0.219 | 0.231 | 0.170 | **0.372** | 0.136 |
| **Input: DP abstracts** | | | | | |
| DP–CT nuggets | **0.321** | 0.301 | 0.263 | 0.311 | 0.312 |
| **Input: DP full papers** | | | | | |
| DP–CT nuggets | 0.032 | 0.000 | 0.144 | * | **0.280** |

Table 11: Pyramid scores were computed for automatic summaries of QA and DP data. The summaries were evaluated using nuggets drawn from QA citation sentences (QA–CT), QA abstracts (QA–AB), and DP citation sentences (DP–CT). LexRank is computationally intensive and so was not run on the DP-PA dataset, as indicated by * (about 4000 sentences). Highest scores for each input source are shown in bold.

Table 11 gives the pyramid score values of the summaries generated by the four automatic summarizers, evaluated using nuggets drawn from the QA citation sentences, QA abstracts, and DP citation sentences. The table also includes results for the baseline random summaries.

When we used the nuggets from the abstracts set for evaluation, the summaries created from abstracts scored higher than the corresponding summaries created from citations and papers. Further, the best summaries generated from citations outscored the best summaries generated from papers. When we used the nuggets from citation sets for evaluation, the best automatic summaries generated from citations outperform those generated from abstracts and full papers. All these pyramid results demonstrate that citations can contain useful information that is not available in the abstracts or the original papers, and that abstracts can contain useful information that is not available in the citations or full papers.

Among the various automatic summarizers, MASCS performed best at this task, in two cases exceeding the average human performance. Note also that the random summarizer outscored the automatic summarizers in cases where the nuggets were taken from a source different from that used to generate the summary. However, one or two summarizers still tended to do well. This indicates a difficulty in extracting the overlapping summary-amenable information across the two sources.

| | Random | C-LexRank | C-RR | LexRank | MASCS |
|---|---|---|---|---|---|
| **System Performance: ROUGE-2** | | | | | |
| **Input: QA citations** | | | | | |
| QA–CT refs. | 0.116 | **0.170** | 0.095 | 0.135 | **0.170** |
| QA–AB refs. | 0.083 | **0.117** | 0.076 | 0.070 | 0.103 |
| **Input: QA abstracts** | | | | | |
| QA–CT refs. | 0.045 | 0.059 | **0.061** | 0.054 | 0.041 |
| QA–AB refs. | 0.121 | 0.136 | 0.122 | **0.203** | 0.134 |
| **Input: QA full papers** | | | | | |
| QA–CT refs. | 0.030 | 0.036 | 0.036 | **0.282** | 0.040 |
| QA–AB refs. | 0.046 | 0.059 | 0.050 | **0.105** | 0.075 |
| **Input: DP citations** | | | | | |
| DP–CT refs. | 0.107 | **0.132** | 0.087 | 0.049 | 0.101 |
| **Input: DP abstracts** | | | | | |
| DP–CT refs. | 0.070 | 0.073 | 0.053 | **0.203** | 0.072 |
| **Input: DP full papers** | | | | | |
| DP–CT refs. | 0.038 | 0.025 | 0.034 | * | **0.046** |

Table 12: ROUGE-2 scores of automatic summaries of QA and DP data. The summaries were evaluated by using human references created from QA citation sentences (QA–CT), QA abstracts (QA–AB), and DP citation sentences (DP–CT). These results are obtained after Jack-knifing the human references so that the values can be compared to those in Table 4. LexRank is computationally intensive and so was not run on the DP full papers set, as indicated by '*' (about 4,000 sentences). Highest scores for each input source are shown in bold.

We then evaluated each of the random summaries and those generated by the four summarization systems against the references. Table 12 lists ROUGE scores of summaries when the manually created 250-word summary of the QA citation sentences, summary of the QA abstracts, and the summary of the DP citation sentences, were used as gold standard.

When we use manually created citation summaries as reference, then the summaries generated from citations obtained significantly better ROUGE scores than the summaries generated from abstracts and full papers ($p < 0.05$) [Result 1]. This confirms that crucial information, amenable to creating a summary and present in citation sentences is not available, or hard to extract, from abstracts and papers alone. Further, the summaries generated from abstracts performed significantly better than those generated from the full papers ($p < 0.05$) [Result 2]. This suggests that abstracts and citations are generally denser in summary-amenable information than full papers.

When we use manually created abstract summaries as reference, then the summaries generated from abstracts obtained significantly better ROUGE scores than the summaries generated from citations and full papers ($p < 0.05$) [Result 3]. Further, and more importantly, the summaries generated from citations performed significantly better than those generated from the full papers ($p < 0.05$) [Result 4]. Again, this suggests that abstracts and citations are richer in summary-amenable information. These results also show that abstracts of papers and citations have some overlapping information (Result 2 and Result 4), but they also have a significant amount of unique summary-amenable information (Result 1 and Result 3).

Among the automatic summarizers, C-LexRank and LexRank perform best. This is unlike the results found through the nugget-evaluation method, where MASCS performed best. This suggests that MASCS is better at identifying more useful nuggets of information, but C-LexRank and LexRank are better at producing unigrams and bigrams expected in a summary. To some extent this may be due to MASCS's compression preprocessing, which breaks large, complex sentences into smaller, finer-grained units of content that correspond better to the amount of content in a nugget.

## 7. Conclusion

In this paper, we investigated the usefulness of directly summarizing citation sentences (set of sentences that cite a paper) in the automatic creation of technical summaries. We proposed C-LexRank, a graph-based summarization model and generated summaries of 30 single scientific articles selected from 6 different topics in the ACL Anthology Network (AAN). We also generated summaries of a set of Question Answering (QA) and Dependency Parsing (DP) papers, their abstracts, and their citation sentences using four state-of-the-art summarization systems (C-LexRank, C-RR, LexRank, and MASCS). We then used two different approaches, nugget-based pyramid and ROUGE, to evaluate the summaries. The results from both approaches and all four summarization systems show that both citation sentences and abstracts have unique summary-amenable information. These results also demonstrate that multidocument summarization—especially technical summary creation—benefits considerably from citations.

We next plan to generate summaries using both citation sentences and abstracts together as input. Given the overlapping content of abstracts and citation sentences, discovered in the current study, it is clear that redundancy detection will be an integral component of this future work. Creating readily consumable technical summaries is a hard task, especially when using only raw text and simple summarization techniques. Therefore, we intend to combine these summarization and bibliometric techniques with suitable visualization methods towards the creation of iterative technical survey tools—systems that present summaries and bibliometric links in a visually convenient manner and which incorporate user feedback to produce even better summaries.

Current work on generating topic summaries is focused almost exclusively on extracting diverse factoid-rich summaries. Meanwhile, the fluency of the produced summaries has been mostly ignored. In future work, we plan to employ some post-processing techniques such as reference scope extraction and sentence simplification, as described by Abu-Jbara and Radev (2011), to generate more readable and cohesive summaries.

## 8. Acknowledgments

We would like to thank Ahmed Hassan, Rahul Jha, Pradeep Muthukrishan, and Arzucan Özgür for annotations and Melissa Egan for preliminary developments. We are also grateful to Ben Shneiderman, Judith Klavans, and Jimmy Lin for fruitful discussions and the anonymous reviewers for insightful readings and constructive guidance. The following authors, Vahed Qazvinian, Dragomir R. Radev, Saif M. Mohammad, Bonnie Dorr, David Zajic, and Michael Whidby were supported, in part, by the National Science Foundation under Grant No. IIS-0705832 (iOPENER: Information Organization for PENning Expositions on Research) awarded to the University of Michigan and the University of Maryland. Any opinions, findings, and conclusions or recommendations expressed in this material are those of the authors and do not necessarily reflect the views of the National Science Foundation. The following authors, Michael Whidby and Taesun Moon were supported, in part, by the Intelligence Advanced Research Projects Activity (IARPA) via Department of Interior National Business Center (DoI/NBC) contract number D11PC20153. The U.S. Government is authorized to reproduce and distribute reprints for Governmental purposes not withstanding any copyright annotation thereon. Disclaimer: The views and conclusions contained herein are those of the authors and should not be interpreted as necessarily representing the official policies or endorsements, either expressed or implied, of IARPA, DoI/NBC, or the U.S. Government.